# Small-Data Machine Learning Uncovers Decoupled Control Mechanisms of Crystallinity and Surface Morphology in β-Ga₂O₃ Epitaxy

*Min Peng[†,▽], Yuanjun Tang[†,▽], Dianmeng Dong[†], Yang Zhang[†,¶], Cheng Wang[†], Shulin Jiao[†], Xiaotong Ma[†], Shichao Zhang[†], Jingchen Wang[†], Huiying Wang[†], Yongxin Zhang[†], Huiping Zhu[¶], Yue-Wen Fang[‡, *], Fan Zhang[†, *], and Zhenping Wu[†*]*

† State Key Laboratory of Information Photonics and Optical Communications & School of Physical Science and Technology, Beijing University of Posts and Telecommunications, Beijing 100876, P. R. China

‡ Centro de Física de Materiales (CFM-MPC), CSIC-UPV/EHU, Manuel de Lardizabal Pasealekua 5, Donostia/San, Sebastián 20018, Spain

¶ Institute of Microelectronics and Key Laboratory of Science and Technology on Silicon Devices, Chinese Academy of Sciences, Beijing 100029, China

**Abstract：** The ultrawide-bandgap semiconductor β-Ga₂O₃ holds exceptional promise for next-generation power electronics and deep-ultraviolet optoelectronics, yet its widespread application is hindered by the lack of cost-effective, high-quality heteroepitaxial thin films. Here, we demonstrate an interpretable machine learning framework that efficiently navigates the complex, multiparameter process space of pulsed laser deposition (PLD) to achieve high-crystallinity $\beta$-Ga$_2$O$_3$ epitaxy on c-plane sapphire. By systematically benchmarking nine regression algorithms under limited experimental data conditions, we identify quadratic polynomial ridge regression as the optimal surrogate model, which combines predictive accuracy ($R^2 \approx 0.86$) with full physical transparency through explicit analytical coefficients. Coupling this model with SHAP (SHapley Additive exPlanations) analysis and iterative experimental design, we construct a closed-loop optimization workflow that progressively refines the process-performance landscape over only three experimental rounds. This data-efficient strategy reduces the X-ray rocking curve (RC) full-width at half-maximum (FWHM) by 70% from >3° to 0.92°, which is the best reported value for PLD-grown $\beta$-Ga$_2$O$_3$ on sapphire. Intriguingly, concurrent modeling of surface roughness reveals that crystalline quality and surface morphology are governed by distinct dominant factors: temperature primarily controls bulk crystallinity, whereas oxygen pressure dictates surface kinetics. This decoupled mechanism, quantitatively captured for the first time via feature importance analysis, provides actionable physical insight for independent optimization of structural and morphological properties. Our work establishes a generalizable, resource-efficient paradigm for intelligent process development in oxide epitaxy and beyond.

## 1. Introduction

β-Ga$_2$O$_3$ has emerged as a frontrunner among ultrawide-bandgap semiconductors, featuring a bandgap of 4.8 - 4.9 eV, a theoretical breakdown field of 8 MV·cm$^{-1}$, and a Baliga figure of merit that substantially surpasses those of SiC and GaN.[1,2] These attributes render it highly attractive for high-power switching devices, solar-blind photodetectors, and harsh-environment electronics.[3–9] Although melt-grown bulk crystals with low defect densities are now available,[10] the limited size, high cost, and poor thermal conductivity of native substrates motivate the development of high-quality heteroepitaxial thin films on foreign substrates such as sapphire, which is a prerequisite for cost-effective device integration and heterostructure engineering.[2,11–13]

Among the various deposition techniques explored for β-Ga$_2$O$_3$, including molecular beam epitaxy (MBE), metal-organic chemical vapor deposition (MOCVD), halide vapor phase epitaxy (HVPE), and sputtering,[11,14–17] pulsed laser deposition (PLD) offers a unique combination of precise stoichiometry transfer, wide process flexibility, and straightforward doping control.[18] However, PLD growth is governed by strongly coupled parameters (substrate temperature, oxygen pressure, laser fluence, etc.),[14] and the conventional trial-and-error approach to process optimization is both time-consuming and resource-intensive, often failing to locate the narrow window for high-quality epitaxy within a limited experimental budget.

Machine learning (ML) has recently emerged as a powerful paradigm for accelerating materials discovery and process optimization in both experimental fabrication and characterization.[19–23] Among recent progress, ML-assisted approaches have demonstrated success in the epitaxial growth of thin films across diverse material systems. Wakabayashi et al. identified optimal combinations of temperature and oxygen pressure for MBE growth of SrRuO$_3$ using Bayesian surrogate models;[24] Ohkubo et al. realized closed-loop optimization of TiN epitaxy by embedding ML predictions directly into the growth controller;[25] By combining ML with in-situ photoluminescence, Shen et al. achieved real-time feedback control of InAs/GaAs quantum dot growth.[26] More complex multimodal frameworks have also been developed for GaSe and transition-metal dichalcogenides, integrating structural, morphological, and spectroscopic descriptors into unified latent spaces.[27,28] Besides these materials, ML has demonstrated its power in optimizing the growth process of Ga$_2$O$_3$. Specifically, ML approach based on random forest algorithm has been used to capture the nonlinear dependencies among various growth parameters of metal-organic vapor phase epitaxy (MOVPE) and optimize the growth rate of β-Ga$_2$O$_3$.[29] More recently, ML-guided MOCVD enabled high-rate (1.2 μm h$^{-1}$) step-flow β-Ga$_2$O$_3$ homoepitaxy, addressing the traditional trade-off between growth rate and crystalline quality. The resulting epilayers display atomically smooth surfaces with record-low roughness (0.121 nm) alongside exceptional electrical properties, making an important step toward the device-grade β-Ga$_2$O$_3$.[30]

While most existing studies prioritize predictive accuracy, they offer limited physical interpretability, leaving the underlying process–structure relationships as a black box. Such a black box without clearly interpretable physics may lead to risk overfitting and extrapolation instability, especially in the epitaxy research with limited size in data. For thin film epitaxy, lattice matching is fundamental to achieving high-quality heteroepitaxy on this substrate. Sapphire is arguably the most accessible and versatile platform for gallium oxide heteroepitaxy[31]. Yet its integration with systematic ML-assisted PLD process optimization remains largely unexplored. Here, we address these challenges by developing an interpretable and closed-loop ML framework tailored for process optimization under limited experimental data, applied to β-$Ga_2O_3$ epitaxy on c-plane sapphire via PLD. **Figure 1** illustrates our workflow, which integrates four iteratively connected modules: (i) Experiment & Characterization, where films are deposited under varied temperature–oxygen pressure conditions and characterized by X-ray diffraction (XRD) and atomic force microscopy (AFM) to extract crystalline quality (i.e. FWHM) and surface roughness ($R_q$, $R_a$) metrics; (ii) Data & ML Modeling, where the accumulated process–property data are used to train and benchmark surrogate regression models; (iii) Interpretability & Visualization, where SHAP analysis and response surface mapping reveal feature importance and identify the predicted optimal region; and (iv) Iteration & Optimization, where model-guided experimental design selects the next batch of growth conditions, closing the loop for progressive refinement. The atomic alignment at the β-$Ga_2O_3$/sapphire interface is schematically illustrated in **Figure 2a**, where the well-matched lattices promise the high-quality epitaxial growth. Through three iterative rounds of epitaxy, we reduce the RC FWHM by 70% from >3° to 0.92°, which is the best reported value for PLD-grown β-$Ga_2O_3$ on sapphire[32]. Beyond optimization, comparative SHAP analysis of FWHM and surface roughness uncovers a decoupled control mechanism: temperature dominates crystalline quality whereas oxygen pressure governs surface morphology. This finding, quantitatively demonstrated here for the first time, provides actionable physical insight for independent tuning of structural and morphological properties in β-Ga$_2$O$_3$ epitaxy and, more broadly, establishes a resource-efficient, interpretable ML paradigm generalizable to other oxide thin-film systems.

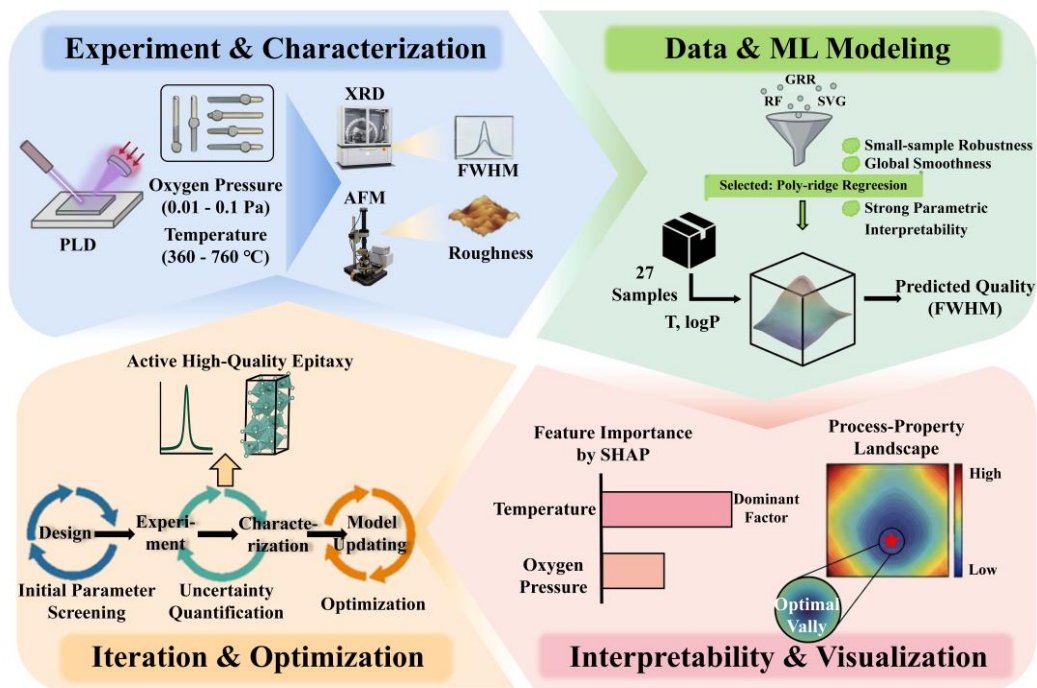

**Figure 1. Schematic of the closed-loop machine learning framework for β-Ga$_2$O$_3$ epitaxial process optimization.** The workflow integrates four modules: (i) Experiment & Characterization: PLD growth under varied temperature and oxygen pressure conditions, followed by XRD and AFM analysis; (ii) Data & ML Modeling: training and benchmarking of surrogate regression models; (iii) Interpretability & Visualization: SHAP-based feature importance analysis and response surface mapping; (iv) Iteration

& Optimization: model-guided selection of next-round experimental conditions, closing the loop for progressive refinement toward optimal crystalline quality.

## 2. Results and discussion

### Phase Diagram and Crystallization Behavior of PLD-Grown $\beta$-$Ga_2O_3$

Following the workflow outlined in **Figure 1**, we first established the crystallization behavior of $\beta$-$Ga_2O_3$ thin films across a broad range of temperature and oxygen pressure conditions (360 – 760 °C, 0.01 – 1 Pa). **Figures S1a–c** present θ-2θ XRD patterns at three representative oxygen pressures. The results reveal a systematic interplay between temperature and oxygen pressure in governing phase formation. At low oxygen pressure (0.01 Pa), films deposited below 510 °C exhibit only broad diffraction envelopes characteristic of an amorphous phase. Above 510 °C, sharp (−201) reflections emerge without detectable secondary phases, indicating well-oriented epitaxial growth. At intermediate oxygen pressure (0.1 Pa), the onset of crystallization shifts downward to ~460 °C, suggesting that enhanced oxygen activity promotes nucleation at lower thermal budgets. At high oxygen pressure (1 Pa), the phase evolution becomes more complex where films transition from amorphous (360 °C) to polycrystalline (410–460 °C) before recovering single-phase epitaxy above 510 °C. This non-monotonic behavior reflects competition between oxygen-enhanced nucleation density and thermally activated adatom mobility, which is a common phenomenon of PLD kinetics in oxide systems.[33]

These observations are summarized in the temperature–oxygen pressure phase diagram (**Figure 2b**), which delineates three distinct regions: amorphous, polycrystalline, and epitaxial single crystal. The high-quality epitaxial window occupies a relatively narrow zone at intermediate temperatures and oxygen pressures, consistent with prior reports on β-$Ga_2O_3$ grown by MBE and MOCVD.[2,11] This phase diagram serves as both a qualitative guide and a ground truth for subsequent ML predictions.

Quantitative assessment of crystalline quality was performed through the X-ray RC measurements of the (−201) reflection. **Figure 2c** summarizes the extracted FWHM as a function of temperature at each oxygen pressure, and **Figures S1d–f** show the representative RCs. These data are also summarized in **Table S1**. At 0.01 Pa, FWHM decreases from 1.92° at 510 °C to a minimum of 1.34° at 610 °C before increasing again at higher temperatures, indicating an optimal thermal window beyond which interface reactions or thermal stress degrade crystallinity.[31,34,35] At 0.1 Pa, the minimum FWHM of 1.18° is achieved at 660 °C. At 1 Pa, FWHM decreases monotonically from 3.89° at 410 °C to 1.38° at 760 °C without reaching a plateau, suggesting that even higher temperatures might yield further improvement under oxygen-rich conditions. These nonlinear oxygen pressure-dependent trends underscore the complex relationship in process–quality, and motivate our exploration using ML-based surrogate modeling.

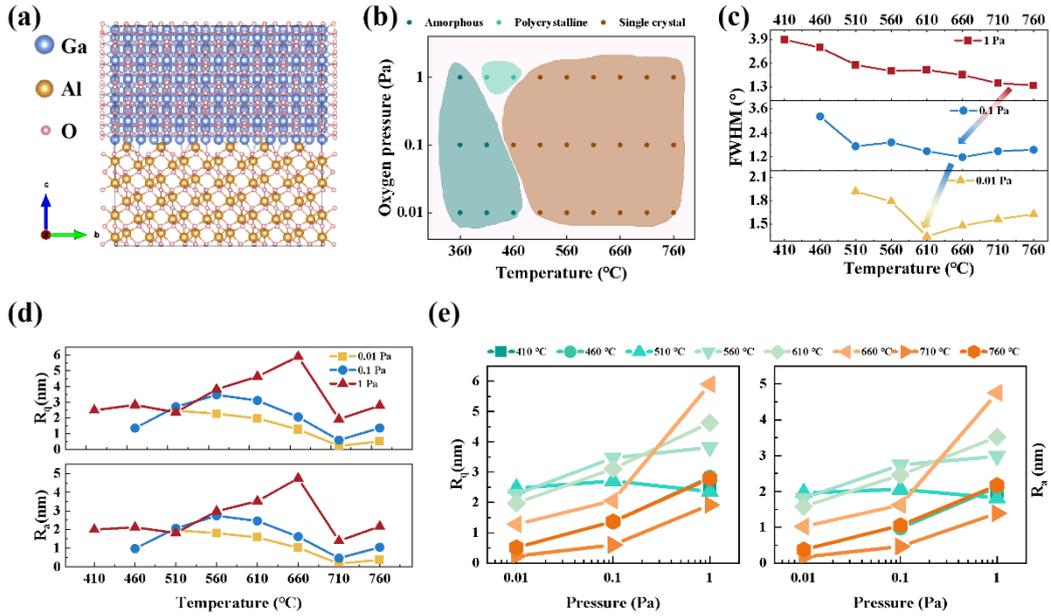

**Figure 2. Structural characterization, Surface Roughness Characterization and phase diagram of PLD-grown β-Ga₂O₃ films on sapphire. (a)** β-Ga₂O₃(-201)/α-Al₂O₃(0001) Interface Atomic Model. **(b)** Temperature–oxygen pressure phase diagram constructed from XRD results, delineating amorphous (dark green), polycrystalline (pale green), and high-quality epitaxial (orange) regions. **(c)** FWHM of the (−201) rocking curve as a function of substrate temperature at different oxygen pressures, revealing the nonlinear, pressure-dependent optimization landscape. **(d)** $R_q$, $R_a$ as a function of substrate temperature at different oxygen pressures, **(e)** $R_q$, $R_a$ as a function of oxygen pressure at different substrate temperatures, revealing the nonlinear, parameter-dependent optimization landscape.

### Surrogate Model Selection for Small-Sample Regression

To select a model architecture that balances predictive accuracy with generalizability and interpretability is very crucial in ML-assisted process optimization. This is particularly important to epitaxy research in which the size in experimental data is very limited. To address this challenge, we benchmarked nine widely used regression algorithms including gradient boosting regression (GBR), quadratic polynomial ridge regression (Ridge-Poly2), kernel ridge regression, random forest, k-nearest neighbors, Bayesian ridge, elastic net, radial basis function support vector regression, and Gaussian process regression with RBF kernel.[19,25,36]

Each model was evaluated using repeated random train–test splits, with performance quantified by root-mean-square error (RMSE), mean absolute error (MAE), and coefficient of determination ($R^2$). **Figure 3a** summarizes the statistical results. GBR achieved the lowest point-estimate errors, while its piecewise-constant response surface exhibited spurious peaks and discontinuities in data-sparse regions, which is a well-known limitation of tree-based ensembles that compromises extrapolation reliability.

In contrast, Ridge-Poly2 offered a compelling trade-off: its RMSE and MAE were only marginally higher than GBR, while its $R^2$ approached 0.8. Crucially, Ridge-Poly2 yields an explicit analytical form, which is a second-order polynomial in temperature and log-pressure with $L_2$ regularization, and its coefficients directly encode the direction and magnitude of each parameter's influence on FWHM. This transparency enables immediate physical interpretation: positive coefficients indicate factors that increase FWHM (degrade quality), while interaction terms reveal synergistic or antagonistic coupling. Furthermore, the smooth quadratic response surface avoids artificial structures in unexplored regions, providing stable extrapolation essential for guiding experiments into uncharted process space. Based on these considerations, we selected Ridge-Poly2 as the surrogate model for all subsequent optimization.

**Iterative Closed-Loop Optimization**

Rather than exhaustively sampling a broad range of temperature and oxygen pressure conditions, we adopted an iterative experimental design strategy that leverages ML predictions to prioritize high-information regions (**Figure 1**). This closed-loop workflow proceeds in three stages: (i) initial coarse sampling, (ii) uncertainty and residual-guided refinement, and (iii) targeted exploitation of the predicted optimum.

Round 1 (Initial Simulation): The first batch of experiments followed a uniform grid spanning the accessible temperature–oxygen pressure range. This dataset established the initial phase diagram (**Figure 3b**) and trained the first-generation surrogate model. As shown in **Figures 3e, h**, the model captured the overall trend but exhibited limited precision ($R^2 \approx 0.7$), with substantial scatter between predicted and measured FWHM values.

Round 2 (Optimized Iteration): Guided by the Round 1 model, we identified regions of high prediction uncertainty (sparse sampling) and large residuals (model–experiment mismatch). Additional experiments (**Figure 3c**) were strategically placed in these regions. Retraining on the augmented dataset improved $R^2$ to ~0.8 (**Figures 3f, i**).

Round 3 (Final Iteration): With confidence in the model's predictive landscape, Round 3 focused on fine-sampling the predicted low-FWHM "sweet spot." Although fewer samples were added (**Figure 3d**), they were concentrated near the response surface minimum. The final model achieved $R^2 \approx 0.86$ (**Figures 3g, j**), with predicted and measured values tightly clustered along the diagonal. The XRD patterns and RCs for the samples added in the latter two iteration rounds are provided in **Figures S2, 3**.

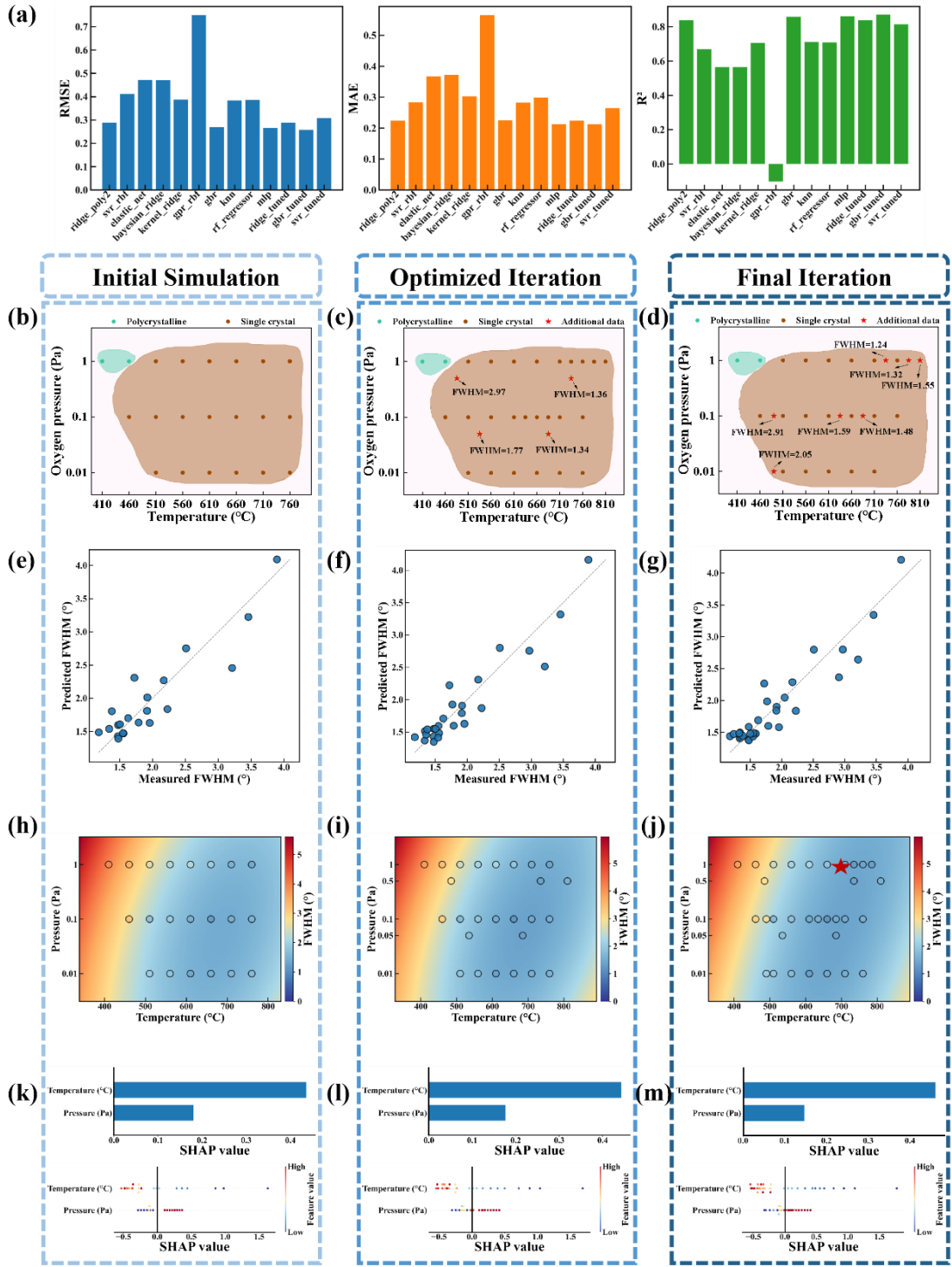

**Figure 3. Evolution of model performance, response surface, and feature importance across three optimization rounds via Surrogate model benchmarking and iterative data augmentation. (a)** Comparison of nine regression algorithms evaluated by RMSE, MAE, and $R^2$ under repeated random train–test splits. Error bars

represent standard deviations. Quadratic polynomial ridge regression (Ridge-Poly2) was selected based on its balance of predictive accuracy and physical interpretability. Subsequent panels illustrate the evolution of the iterative optimization strategy: Initial Simulation (Round 1), Optimized Iteration (Round 2), and Final Iteration (Round 3). Specifically, **(b-d)** Temperature–oxygen pressure coordinates of samples, illustrating the progression from uncertainty-guided refinement to exploitation of the predicted optimal region. **(e-g)** Predicted versus measured FWHM scatter plots, showing progressive improvement in model accuracy ($R^2$ from ~0.7 to ~0.86). Dashed lines indicate ideal 1:1 correspondence. **(h-j)** FWHM response surfaces in the temperature–$\log_{10}p(O_2)$ parameter space, with experimental data points overlaid. The predicted optimal region (dark blue valley) sharpens and converges with successive iterations. **(k-m)** SHAP analysis results: mean absolute SHAP values indicating global feature importance; SHAP dependence plots showing the directional influence of temperature and oxygen pressure on predicted FWHM. The progressive rebalancing of feature importance reflects improved model capture of the underlying process–property relationships.

Across three iterations, the global feature importance derived from SHAP analysis consistently identifies temperature as the dominant feature influencing FWHM **(Figures 3k, l, m)**. The SHAP-based importance of temperature remained significantly higher than that of oxygen pressure throughout all rounds. As the dataset expanded, the relative importance of temperature further intensified, while the influence of oxygen pressure declined in the third-round model, indicating that the temperature is the dominant factor governing FWHM. The converged response surface **(Figure 3j)** displays a well-defined valley corresponding to the optimal process window. Verification experiments within this valley yielded a minimum FWHM of 0.92° **(Figure 4a)**, which reduces >70% from the initial >3° values observed in Round 1. This result surpasses the previously reported best value of 1.1° for PLD-grown β-Ga$_2$O$_3$ on sapphire and is competitive with optimized MBE and MOCVD films on the same substrate.[15,37–39] **Figure 4b** places our result in the context of literature data across different growth techniques[15,32,37–48], confirming that interpretable, small-data ML can match or exceed outcomes achieved by far more resource-intensive conventional optimization.

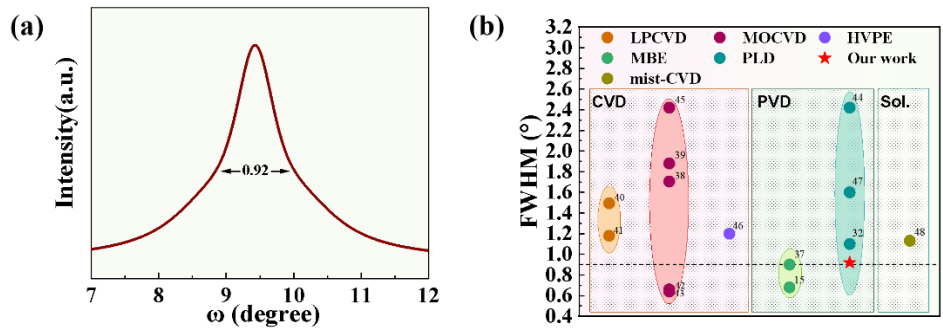

**Figure 4. Optimized epitaxial quality and comparison with literature. (a)** X-ray rocking curve of the (−201) reflection for the sample grown under ML-predicted optimal conditions, yielding FWHM = 0.92°. **(b)** Comparison of FWHM values for β-Ga$_2$O$_3$ films on sapphire across different growth techniques (e.g., PLD, MBE, MOCVD, HVPE) reported in the literature. The result from this work (red star) represents the best reported value for PLD-grown β-Ga$_2$O$_3$ on sapphire.

The efficiency gain is substantial: the entire optimization from initial exploration to final verification required only ~30 samples partitioned across three experimental rounds. In contrast, a conventional full-factorial design covering the same parameter space at comparable resolution would demand >100 experiments. This five-fold reduction in experimental burden demonstrates the practical value of ML-guided iterative design for accelerating epitaxial process development.

**Surface Morphology and Growth Mode Evolution of PLD-Grown $\beta$-Ga$_2$O$_3$**

The surface roughness evolution of the $\beta$-Ga$_2$O$_3$ films under varying deposition conditions is systematically summarized in **Figures 2d, e**. Representative AFM images for each sample are provided in the Supporting Information (**Figure S4**), which visually corroborate the quantitative trends discussed below. Data of both the root mean square roughness $R_q$ and the arithmetic average roughness $R_a$ are also summarized in **Tables S2, 3**.

**Figure 2d** illustrates the variation of surface roughness ($R_q$ and $R_a$) of $\beta$-Ga$_2$O$_3$ films with substrate temperature under three different oxygen pressures: 0.01 Pa, 0.1 Pa, and 1 Pa. Overall, the roughness exhibits a nonlinear evolution with increasing temperature, and its behavior depends significantly on the oxygen pressure. At 0.01 Pa, the roughness decreases almost monotonically with temperature, reaching its minimum at 710 °C ($R_q$ = 0.232 nm, $R_a$ = 0.17 nm). Under 0.1 Pa, the roughness first increases and then decreases with temperature, also attaining its lowest value at 710 °C ($R_q$ = 0.602 nm, $R_a$ = 0.465 nm). At 1 Pa, the temperature dependence of roughness is more complex, yet it still reaches its minimum at 710 °C under this pressure condition ($R_q$ = 1.92 nm, $R_a$ = 1.39 nm). Furthermore, at a given temperature, lower oxygen pressures generally correspond to lower surface roughness, which is more directly visualized in **Figure 2e**, which plots the roughness as a function of oxygen pressure.[49]

Collectively, the data from **Figures 2d, e** reveal that optimal surface smoothness is accessible only within a synergistic process window combining relatively high temperature and low-to-moderate oxygen pressure. This nonlinear interdependence further underscores the complexity of the process–quality relationship and motivates the adoption of ML-based surrogate modeling for global process optimization.

**Decoupled Dominant Factors for Crystallinity versus Surface Morphology**

Beyond process optimization, this framework offers mechanistic insights through a comparative analysis of different quality metrics. In parallel with FWHM modeling, we constructed surrogate models for surface roughness (Ra and Rq) measured via AFM. Unlike FWHM, which was adequately described by a second-order model (**Figure S5**), we found that low-order polynomials failed to effectively fit the roughness data; instead, a fourth-order polynomial regression was required to capture its complex evolutionary characteristics. The optimized models exhibited satisfactory accuracy: $R^2$ = 0.7740 (RMSE = 0.5690 nm, MAE = 0.3907 nm) for Ra, and: $R^2$ = 0.7860 (RMSE = 0.6914 nm, MAE = 0.4893 nm) for Rq (**Figure 5a**). This necessity for higher-order modeling intuitively reflects the high degree of nonlinearity and complex parameter coupling governing surface morphology evolution.

SHAP analysis (**Figure 5c**) reveals a significant shift in the distribution of feature importance. For FWHM, the SHAP value of temperature is substantially higher than that of oxygen pressure, indicating an overwhelming dominance. This aligns with conventional expectations for epitaxial growth, where bulk crystallinity is primarily controlled by the thermal condition, which dictates atom diffusion lengths and defect annihilation kinetics. In contrast, the high-precision roughness models present a distinctly different picture: although the gap between temperature and oxygen pressure SHAP values narrows (indicating that both play significant roles under fourth-order coupling), a critical reversal in relative ranking occurs, with oxygen pressure emerging as the primary factor influencing surface Ra. This suggests that surface morphology is more sensitive to oxygen-related kinetics, including plume scattering, reactive oxygen species concentration, and adsorption–desorption equilibria, rather than being governed primarily by the thermally activated factors that dominate bulk ordering.

This high-order nonlinearity manifests on the response surface (**Figure 5b**) as two distinct optimal regions (minima) within the roughness landscape, in sharp contrast to the single "sweet spot" observed for FWHM. This difference in process window structure offers flexibility for application-oriented process selection: for devices sensitive to bulk defects, one can prioritize the FWHM optimal domain; whereas for those sensitive to interface scattering, the AFM optimal minima can be targeted. Crucially, the incomplete overlap of these landscapes implies the existence of a distinct Pareto frontier for multi-objective optimization. This allows researchers to navigate the parameter space to balance trade-offs based on specific device tolerances for crystalline quality versus surface morphology, rather than pursuing a single global optimum.

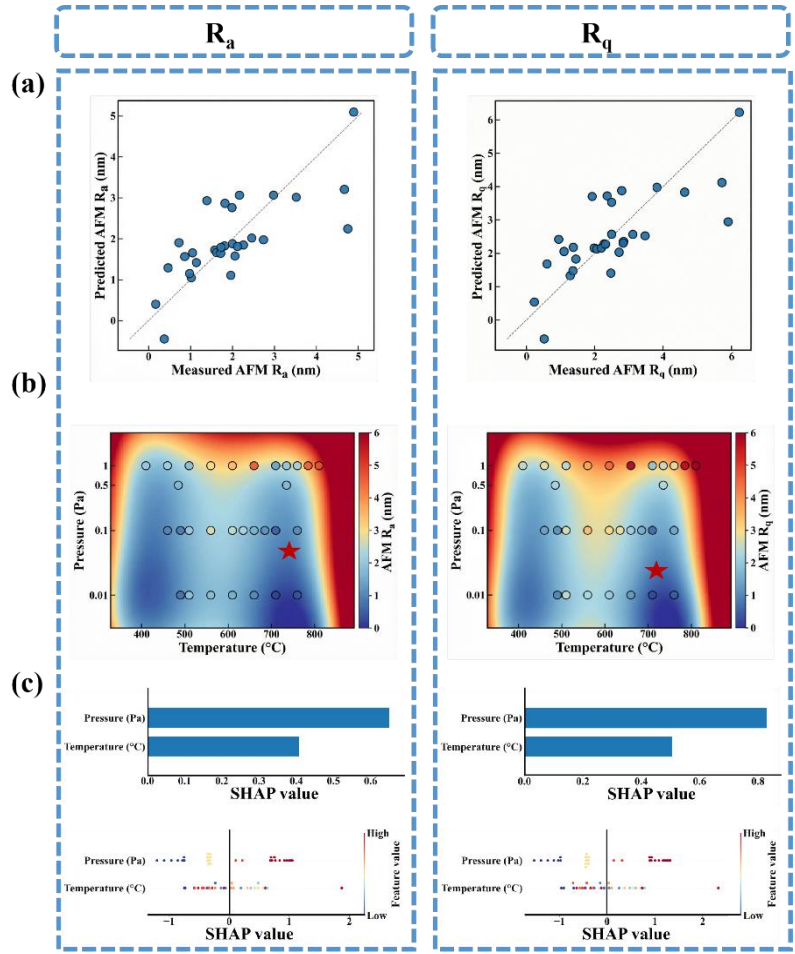

**Figure 5. Surface roughness fourth-order model performance, response surfaces and feature importance. (a)** Predicted versus measured $R_a$, $R_q$ scatter plots, the model achieved $R^2 = 0.7740$ ($R_a$) and $0.7860$ ($R_q$). **(b)** $R_a$, $R_q$ response surfaces in the temperature $-\log_{10}p(O_2)$ parameter space, with experimental data points overlaid. Two distinct optimal regions for $R_a$ and $R_q$ offers section for multi-objective optimization. **(c)** SHAP analysis results and SHAP dependence plots showing the directional influence of temperature and oxygen pressure on predicted FWHM. Oxygen pressure emerges as the primary factor influencing surface $R_a$, $R_q$.

## 3. Conclusions

In summary, we have developed an interpretable machine learning framework for efficient process optimization of β-Ga$_2$O$_3$ heteroepitaxy on sapphire via pulsed laser deposition. By systematically benchmarking regression algorithms and selecting quadratic polynomial ridge regression as the optimal surrogate model, we established a closed-loop workflow integrating SHAP-based feature analysis with iterative experimental design. This data-efficient strategy reduced the X-ray rocking curve FWHM by >70% from >3° to 0.92°, within only three optimization rounds using ~30 samples, representing the best reported value for PLD-grown β-Ga$_2$O$_3$ on sapphire. Comparative analysis further revealed that crystalline quality and surface morphology are governed by distinct dominant factors, providing mechanistic insight for independent property optimization. The framework demonstrated here requires no specialized equipment and is readily transferable to other oxide epitaxial systems, offering a resource-efficient paradigm for intelligent thin-film process development.

## 4. Experimental sections

### *Film Growth:*

On c-plane (0001) sapphire substrates, β-Ga$_2$O$_3$ thin films were deposited by pulsed laser deposition using a KrF excimer laser ($\lambda = 248$ nm). A polycrystalline Ga$_2$O$_3$ target (99.99% purity) was used as the source material, positioned 70 mm from the substrate. The laser repetition rate was fixed at 2 Hz with a fluence of ~100 mJ·cm$^{-2}$ at the target surface. Prior to deposition, the chamber was evacuated to a base vacuum below $1 \times 10^{-4}$ Pa. The substrate temperature was varied from 360 to 760 °C, and high-purity oxygen was introduced through a mass flow controller to achieve working pressures ranging from 0.01 to 1 Pa. Each deposition consisted of 6000 laser pulses. The target was pre-ablated for several minutes before each deposition to stabilize the plume and was continuously rotated during growth to ensure uniform erosion.

### *Structural and Morphological Characterization*

Crystalline phase and orientation were determined by high-resolution XRD using Cu Kα radiation. θ-2θ scans were performed to identify phase purity and preferential orientation. RC measurements of the (−201) reflection were conducted to quantify crystalline quality; the full-width at half-maximum (FWHM) was extracted by Gaussian fitting. For samples exhibiting multipeak or shoulder features, an equivalent single-peak fitting procedure was applied to ensure data consistency across the sample set.

Surface morphology was characterized by atomic force microscopy operating in tapping mode. Images were acquired over $10 \times 10$ μm$^2$ areas, from which the arithmetic mean roughness ($R_a$) and root-mean-square roughness ($R_q$) were extracted.

### *Machine Learning Modeling*

Each PLD experiment was treated as a single data point. The input features were the substrate temperature (T) and the base-10 logarithm of oxygen pressure, $\log_{10} p(O_2)$, while the FWHM served as the output target. Nine regression algorithms were evaluated: gradient boosting regression, quadratic polynomial ridge regression (Ridge-Poly2), kernel ridge regression, random forest, k-nearest neighbors, Bayesian ridge, elastic net, support vector regression with radial basis function kernel, and Gaussian process regression with RBF kernel. All models were implemented using scikit-learn (v1.2). Hyperparameters were tuned via grid search with cross-validation.

Model performance was assessed using repeated random train–test splits (80:20 ratio, 50 repetitions) to obtain robust estimates of root-mean-square error (RMSE), mean absolute error (MAE), and coefficient of determination ($R^2$). Based on the trade-off between predictive accuracy, extrapolation stability, and physical interpretability, Ridge-Poly2 was selected as the surrogate model for process optimization.

### *Interpretability Analysis*

Feature contributions were quantified using SHAP (SHapley Additive exPlanations), which decomposes each prediction into additive contributions from individual features based on cooperative game theory. Mean absolute SHAP values were computed to rank global feature importance, while SHAP dependence plots were used to visualize how each feature influences predictions across the sample set. All SHAP analyses were performed using the SHAP Python library (v0.41).

### *Iterative Experimental Design*

Process optimization followed a three-round iterative strategy. In Round 1, experiments were distributed on a uniform grid spanning the accessible temperature–pressure space to establish an initial phase diagram and train the first-generation surrogate model. In Round 2, experiments were placed in regions of high model uncertainty or large prediction residuals to improve model fidelity. In Round 3, experiments were concentrated near the predicted optimum to verify and refine the identified process window. After each round, newly acquired data were incorporated into the training set and the model was retrained to update the response surface.

## ASSOCIATED CONTENT

**Supporting Information.**

Additional characterization data, including Figures S1-S5 and Tables S1-S3.

## AUTHOR INFORMATION

**Corresponding Author**

Yue-Wen Fang - Email: yuewen.fang@csic.es

Fan Zhang - Email: fzhang@bupt.edu.cn

Zhenping Wu - Email: zhenpingwu@bupt.edu.cn

**Author Contributions**

∇Ming Peng and Yuanjun Tang contributed equally. The manuscript was written through contributions of all authors. All authors have given approval to the final version of the manuscript.

**Notes**

The authors declare no competing financial interest.

**Acknowledgments**

This work was supported by the National Natural Science Foundation of China (No. 12474065), the Beijing Nature Science Foundation (Z250006), the Fund of State Key

Laboratory of Information Photonics and Optical Communications (IPOC2025ZR05, IPOC2025ZJ06, IPOC2024ZT01), and the Fundamental Research Funds for the Central Universities (BUPT, 2023ZCJH1, 530524002). Y. T. acknowledges the China Scholarship Council (Grant No. 202506470039). Y.-W.F. thanks the Extraordinary Grant of CSIC (No. 2025ICT122).

**Data availability**

All data supporting the results of this study are available in the manuscript or the supplementary information. Additional data is available from the corresponding author upon request.